\documentclass[12pt]{article}   
\usepackage{graphicx}

\begin{document}

\title{Quantitative scattering of melanin solutions}

\author{Jennifer Riesz$^\ddag$ \footnote{Corresponding author. Tel.: +61 7 3365 3406; Fax: +61 7 3365 1242 E-mail address: riesz@physics.uq.edu.au}, Joel Gilmore$^\dagger$ and Paul Meredith$^\ddag$}
\date{\today}
\maketitle
$^\ddag$ Soft Condensed Matter Group and Centre for Biophotonics and Laser Science, $^\dagger$ Condensed Matter Theory Group, Physics Department, University of Queensland, St. Lucia, Brisbane, QLD 4072, Australia

\begin{abstract}
The optical scattering coefficient of a dilute, well solubilised eumelanin solution has been accurately measured as a function of incident wavelength, and found to contribute less than $6\%$ of the total optical attenuation between 210 and 325nm.  At longer wavelengths (325nm to 800nm) the scattering was less than the minimum sensitivity of our instrument.  This indicates that UV and visible optical density spectra can be interpreted as true absorption with a high degree of confidence.  The scattering coefficient vs wavelength was found to be consistent with Rayleigh Theory for a particle radius of $38\pm1$nm.
\end{abstract}


\section{Introduction}
Melanin is a biological pigment found in the skin, hair and eyes of many species, including humans.  It is thought to be a photoprotectant, but paradoxically has also been implicated in the chain of events that lead to melanoma skin cancer \cite{Hill95, Nofsinger99, Menon97}.  Of the two types found in human skin (eumelanin and pheomelanin) eumelanin is the most common, and the most extensively studied.  Eumelanin is known to be a macromolecule of DHI (dihydroxyindole) and DHICA (dihydroxyindole-carboxylic acid), but the nature of the secondary structure (i.e. the supramolecular organisation) is not known.

Likely related to its photoprotective role, eumelanin has a broadband absorption spectrum that increases exponentially towards the ultraviolet.  This is a highly unusual feature; most biological pigments exhibit distinct absorption bands.  The origin of the broadband absorption spectrum of eumelanin has long been the topic of scientific debate, which continues to this day.  Galvao and Caldas have used Huckel Theory to attempt to reproduce the broadband shape, with some success \cite{Galvao88, Galvao90I, Galvao90II}.  More recently, Density Functional Theory has been used to predict the optical properties of small eumelanin oliogmers \cite{Stark05, Ilichev03, BolivarMarinez99, Stark03I, Stark03II, Powell05, Powell04}.  This has lead to the theory that eumelanin may in fact be a collection of different small oligomers of varying electronic structure.  The broadband absorption of eumelanin may then be due to the summation of these individual spectra \cite{Stark05, Stark03I, Stark03II}.  This idea was recently extended by the suggestion that the broadband absorption may be due to extreme chemical disorder \cite{Meredith05, Tran05}.

Wolbarsht first suggested that the broadband absorption spectrum may be due to scattering, rather than electronic or physical properties of the eumelanin itself \cite{Wolbarsht81}.  He noted that Rayleigh scattering would reproduce the broadband spectrum, and account for the increase in optical density at short wavelengths.  This has very serious implications; if the measured shape of the absorption spectrum is dominated by scattering then great care must be taken when calculating optical properties.  Despite several studies on the topic, optical scattering remains a significant concern.  The published literature on the scattering of eumelanin solutions is sparse and not cohesive, hence it is useful at this point to briefly review past work.

The importance of optical scattering was noted by Nofsinger and Simon when they discovered that the shape of the eumelanin absorption spectrum is strongly dependant upon the particle size \cite{Nofsinger99, Nofsinger01}.  Since scattering intensity is very strongly dependant upon particle size this could indicate that the optical density of eumelanin is dominated by scattering.  To test this, they conducted photoacoustic measurments which suggested that the measured optical density was not dominated by scattering for wavelengths longer than 400nm for any particle size fraction \cite{Nofsinger99}.  An earlier photoacoustic calorimetry measurement by Forest and Simon similarly suggested that scattering contributes no more than $15\%$ of the total light extinction at 350nm \cite{Forest98}.  Hence Nofsinger and Simon concluded that the observed dependance upon particle size was due to electronic and physical properties of the eumelanin.

Recently, a number of optical emission and excitation studies have been published, which report accurate quantitative measurement of key properties such as the radiative quantum yield as a function of wavelength \cite{Meredith04, Riesz04, NighswanderRempel05A, NighswanderRempel05B, Riesz05CDM}.  Such studies provide valuable insight into energy absorption and dissipation mechanisms, as well as shedding light on the structural question.  These measurements require the assumption that scattering is negligible.  If this is not the case, the scattering coefficient should be measured and subtracted from the optical density to obtain the true absorption.  This was attempted by Krysciak, who directly measured the optical scattering from a dilute eumelanin solution as a function of wavelength \cite{Krysciak85}.  He found scattering to be negligible between 500 and 700nm, but also discovered the puzzling result of `negative scattering' at shorter wavelengths.  He suggested that this was due to multiple scattering events and absorption (which becomes very large at shorter wavelengths) decreasing the measured scattering below the previously measured baseline.  Krysciak's results were non-conclusive, neither confirming nor excluding the prescence of scattering at optical wavelengths.

The following year, Kurtz reported on a theoretical prediction of the relative contributions of scattering and absorption to the optical density of eumelanin \cite{Kurtz86Scat}.  He found that in the Rayleigh regime (particle radii much less than the wavelength) absorption dominated over scattering, whereas for larger particles the two contributed equally.  He emphasised the very strong dependance of scattering on particle size.  The importance of this is experimentally apparent in a 2001 study by Sardar et al.  where scattering and absorption coefficients were measured at four optical wavelengths between 633nm and 476nm \cite{Sardar01}.  They found that scattering far outweighed absorption at all wavelengths, contributing more than $99\%$ of the optical density at 633nm.  This result contradicts all previous studies, and is almost certainly due to the sample preparation, which resulted in what was described as `a brown turbid suspension'.  The authors state that the eumelanin particles were not solubilised and remained a particulate suspension.  Under these conditions, the particle sizes would most likely be much larger than those in the well solubilised, dilute solutions typically used for spectroscopic studies \cite{Meredith04, Riesz04, Nofsinger01, Nofsinger99, NighswanderRempel05A, NighswanderRempel05B}.  

Other studies have attempted to use alternative methods to measure the absorption of eumelanin in the absence of scattering effects.  Caiti et al. used photoacoustic phase angle spectroscopy of powdered melanins in the dry state \cite{Caiti93}.  This technique is insensitive to scattering, and confirmed unambiguously the decrease in the absorption of melanins with increasing wavelength.  Unfortunately, the phase spectra do not correspond by visual inspection to absorption spectra, and interpretation remains difficult.  Therefore, while this study sheds doubt on the Wolbarsht model, it does not allow correction of absorption spectra for scattering effects in a quantitative way.  Similarly, a recent study by Albuquerque et al. used photopyroelectric spectroscopy to measure the optical absorption coefficient of eumelanin in the solid state \cite{Albuquerque05}.  Again, the decrease in absorption with increasing wavelength was confirmed, although a direct comparison with solution measurements could not be made due to the different properties of the system.  Interestingly, a band gap was observed at 1.70eV (730nm), which is possibly hidden in solution spectra by scattering.

A careful study by Vitkin et al. in 1994 gives the most quantitative estimate available of the scattering coefficient of a eumelanin solution \cite{Vitkin94}.  Vitkin et al. conducted photometric measurements with a double integrating sphere system at 580nm and 633nm.  They found that scattering contributed $12\%$ and $13.5\%$ of the total attenuation coefficient at each wavelength respectively.  These values, while small, are enough to introduce significant error in the measurement of the radiative quantum yield and other optical parameters, and should ideally be corrected for.  A measurement of the scattering coefficient as a function of wavelength would allow the subtraction of scattering effects from the optical density spectrum to achieve this.  

If the scattering coefficient as a function of wavelength were known, the shape of the scattering spectrum could be compared with Rayleigh Theory.  As stated earlier, there remains debate as to the secondary structure of eumelanin: heteropolymer or nanoaggregate \cite{Powell04, Powell05}.  This is a most fundamental question, since it influences the interpretation of many other experiments.  Since Rayleigh scattering is strongly dependant upon particle size, these scattering measurements can also be used to determine a fundamental particle size of eumelanin in solution.  Hence we have conducted an integrated scattering measurement as a function of wavelength over the ultraviolet range, where scattering effects should be most significant.

In addition, the solutions used by Vitkin et al. ($0.07\%$ to $0.12\%$ eumelanin by weight) were more concentrated than those best suited to photoluminescence measurements.  The broadband absorption spectrum of eumelanin gives rise to significant reabsorption and inner filter effects at concentrations above $0.0025\%$ by weight \cite{Meredith04, Riesz04}.  Although scattering should scale linearly with concentration it is feasible that there is less aggregation  at lower concentrations, giving rise to less scattering.  Hence we have made a direct measurement of the scattering coefficient at the ideal spectroscopic concentration. 

In this study we:
\begin{enumerate}
\item Measure the integrated scattering from an optical spectroscopy grade eumelanin solution as a function of wavelength from 210nm to 325nm
\item Develop general equations to calculate the scattering in broadband absorbing samples, and apply these to the specific case of a eumelanin solution
\item Show that the measured scattering is consistent with Rayleigh theory, and use this to estimate an approximate particle size
\end{enumerate}

\section{Experimental}

\subsection{Sample Preparation:}
Synthetic eumelanin (dopamelanin) derived from the non-enzymatic oxidation of tyrosine was purchased from Sigma Aldrich, and used without further purification. The powder was solubilized to form a $0.1\%$ solution (by weight) in high purity $18.2M\Omega$ MilliQ de-ionised water.  This stock solution was then diluted to a concentration (by weight) of $0.0025\%$. To aid solubility, the pH of the solution was adjusted to approximately pH11.5 using NaOH, and the solution gently heated with sonication. Under such conditions a pale brown, apparently continuous eumelanin dispersion was produced. This is identical to the sample preparation typically used for spectroscopic analysis \cite{Meredith04, Riesz04}.  This concentration is usually selected since it maximises the weak photoluminescence signal whilst minimising distorting re-absorption and probe beam attenuation effects.

\subsection{Absorption Spectrometry:}
An absorption spectrum between 200nm and 800nm was recorded for the synthetic eumelanin solution using a Perkin Elmer Lambda 40 spectrophotometer. An integration of 2nm, scan speed of 240nm/min and slit width of 3nm bandpass were used. The spectrum was collected using a quartz 1cm square cuvette. Solvent scans (obtained under identical conditions) were used for background correction.

\subsection{Integrated Scattering:}
\begin{figure}
\centering
\includegraphics[width=8cm]{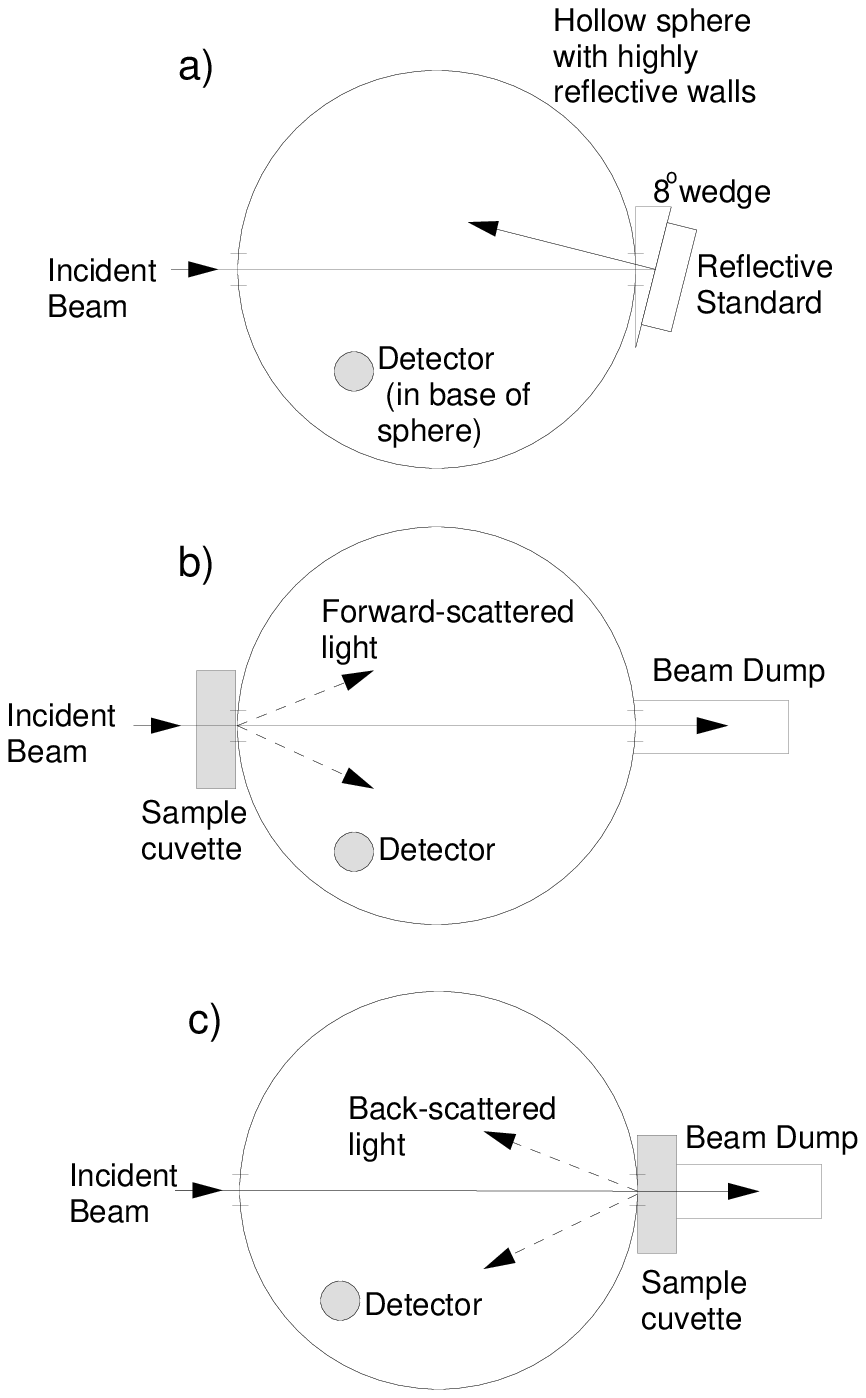}
	\caption{}
	\label{fig:ScatteringApparatus}
\end{figure}

Scattering measurements were made using a Perkin Elmer Lambda 40 spectrophotometer with an integrating sphere attachment (Labsphere RSA-PE-20 Reflectance spectroscopy accessory).  The solution was contained within a 1mm path length quartz cuvette that was placed at the front and back of the sphere as shown in figure \ref{fig:ScatteringApparatus} b) and c) to measure the forwards and backwards integrated scattering respectively. Measurements were taken with a scan speed of 120nm/min, a slit width of 4nm bandpass and 2nm smoothing. Since the scattering intensity was very low each scan was taken five times and averaged.  The $100\%$ reflectance intensity was determined using a labsphere certified reflectance standard, as shown in figure \ref{fig:ScatteringApparatus} a).  The solvent alone was measured in both the front and back positions (figure \ref{fig:ScatteringApparatus} b) and c)) and subtracted after absorption correction (described in the following section). Some light was inevitably lost due to the non-zero size of the beam entry and exit holes in the sphere, and due to the width of the cuvette.  This loss, along with the non-perfect reflectivity of the inside of the sphere was accounted for by the use of the $100\%$ transmission measurement as a standard.  A short path length cuvette (1mm) was used to minimise this loss.


\section{Theory}
Eumelanin solutions have strong, broadband absorbance, and all optical spectroscopic results are therefore affected by re-absorption (attenuation of photoluminescence) and inner filter (attenuation of the incident beam) effects.  Although a narrow cuvette and dilute concentration were used to minimise these effects, it was necessary to perform a careful analysis to account for attenuation of the measured scattering by absorption.  We derive here a general method for correcting for absorption effects in scattering measurements that can be applied to any strongly absorbing solution.  

We define $\alpha_{sf}$ to be the forward scattering coefficient, $\alpha_{sb}$ to be the backward scattering coefficient, and $\alpha_{s}$ to be the total scattering coefficient, such that $\alpha_{sf} + \alpha_{sb} = \alpha_{s}$.  The absorption coefficient is given by $\alpha_a$ and the total attentuation coefficient is given by $\alpha_t$.  We assume that $\alpha_a = \alpha_t - \alpha_s$ (any attenuation not due to scattering is included in the absorption coefficient).  Consider a cuvette of width $d$, with a beam of light incident from the left, as shown in figure \ref{fig:cuvette1}.  By definition, in a small region $dx$ the attenuation of the beam due to each effect (scattering or absorption) is proportional to each $\alpha dx$, and to the intensity of the beam in that region ($I(x)$).  Therefore:
\begin{eqnarray*}
dI(x) &=& -\alpha_{sf}I(x)dx - \alpha_{sb}I(x)dx - \alpha_aI(x)dx \\
I(x) &=&  I_0 e^{-\alpha_t x}
\end{eqnarray*}
which is the familiar Beer-Lambert Law, where $I_0$ is the intensity of light incident upon the cuvette.  Therefore the intensity of light scattered forward ($I_{sf}$) is given by:
\begin{eqnarray}
I_{sf} &=& \int_0^d dI_{sf} \nonumber\\
  &=& \int_0^d \alpha_{sf} \left[I_0 e^{-\alpha_t x}\right]dx \nonumber\\
  &=& \frac{\alpha_{sf}}{\alpha_t}I_0\left(1-e^{-\alpha_td}\right) \label{eq:3}
\end{eqnarray}

\subsection{Correction for Absorption}
\begin{figure}
\centering
\includegraphics[width=6cm]{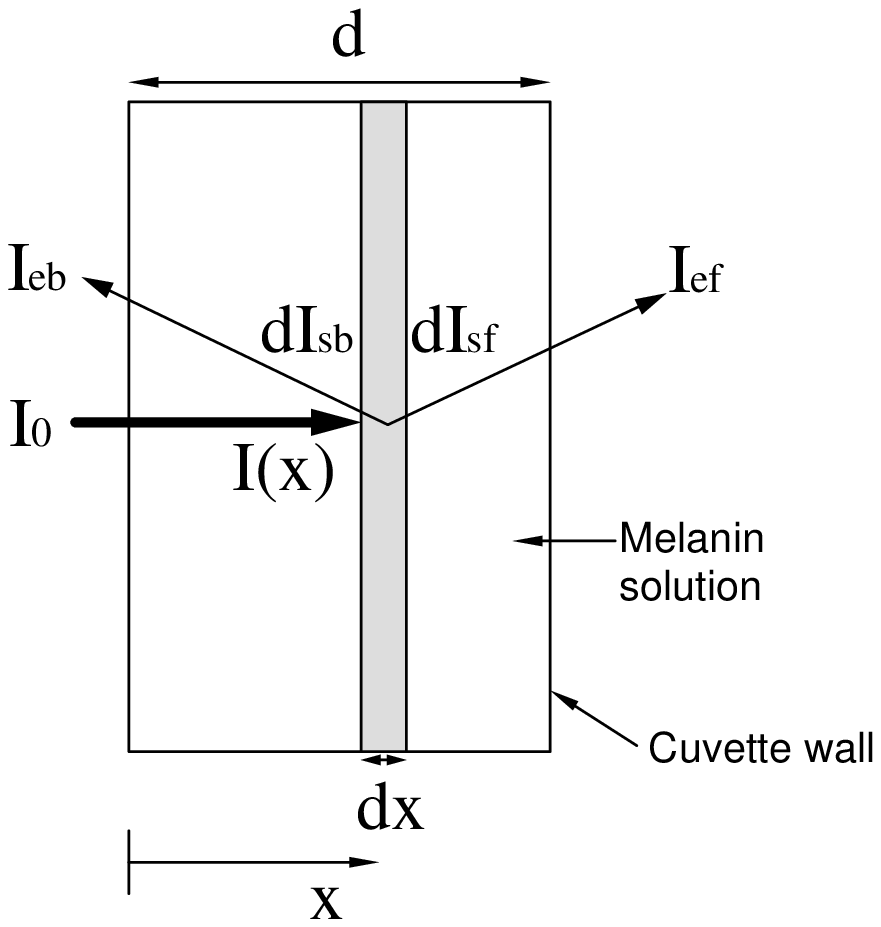}
	\caption{}
	\label{fig:cuvette1}
\end{figure}

We make the geometric approximation that the light scattered in the forward direction will travel a path length of $d-x$ to leave the cuvette (refer to figure \ref{fig:cuvette1}).  As the scattered light travels this distance through the eumelanin solution it will be attenuated by absorption.  We assume that attenuation is only due to absorption here, and not scattering, since multiple scattering events will still be detected.  Let the final intensity emitted forwards from the cuvette (attenuated by absorption) be given by $I_{ef}$.  Using the Beer-Lambert Law:
\begin{eqnarray}
I_{ef} &=& \int_0^d dI_{ef} \nonumber\\
 &=& \int_0^d \left( e^{-\alpha_a(d-x)} dI_{sf} \right)\nonumber\\
 &=& \int_0^d \left( e^{-\alpha_a(d-x)} \alpha_{sf}e^{-\alpha_t x}I_0 dx \right)\nonumber\\
 &=& \frac{\alpha_{sf}}{\alpha_t-\alpha_a}e^{-\alpha_a d}\left[1-e^{-(\alpha_t-\alpha_a)d}\right]I_0 \label{eq:4}
\end{eqnarray}
To determine the amount of light that was originally scattered ($I_{sf}$) from the attenuated intensity that we measure ($I_{ef}$) we combine equations \ref{eq:3} and \ref{eq:4} to eliminate $I_0$:
\begin{equation}
I_{sf} = \frac{\alpha_t-\alpha_a}{\alpha_t}\left(\frac{1-e^{-\alpha_td}}{e^{-\alpha_ad}-e^{-\alpha_td}}\right)I_{ef} - B_f \label{eq:Intensitycorrection}
\end{equation}
where we must subtract off the background signal ($B_f$) which is measured from a blank cuvette (containing solvent only) to remove scattering from the solvent and cuvette walls.  This process can be repeated in a very similar manner for the backwards scattering to find:
\begin{equation}
I_{sb} = \frac{\alpha_t+\alpha_a}{\alpha_t}\left(\frac{1-e^{-\alpha_t d}}{1-e^{-(\alpha_t+\alpha_a)d}}\right)I_{eb}-B_b
\end{equation}
where $I_{sb}$ is the intensity of light scattered backwards, $I_{eb}$ is this intensity attenuated by absorption, and $B_{b}$ is the background scattering in the backwards direction. Note that the different form of the equation is due to the fact that the absorption for back scattering is calculated over a distance $x$ rather than $d-x$, as shown in figure \ref{fig:cuvette1}.

\subsection{Comparison with Experiment}
We must now take into account the actual manner in which the intensity of the scattered light was measured.  We define $S$ to be the light received by the detector as a percentage of the maximum light received with a standard reflector in place of the beamdump (refer to figure \ref{fig:ScatteringApparatus}):
$$S = \frac{I_{recorded}}{I_{max}}\times 100\% $$
Assuming the detector receives a constant fraction of the true scattered light, and $100\%$ of the light is scattered by the standard reflector in the calibration test:
\begin{equation}
S = \frac{I_{scatt}}{I_0}\times 100\% \label{eq:6}
\end{equation}
where $I_{scatt}$ is scattering in either the forwards or backwards direction.  Thus $S$ is the percentage of incident light scattered by the sample.  However, the detected values are affected by absorption.  Let $S_{mf}$ be the scattering signal actually measured (affected by absorption):
$$S_{mf} = \frac{I_{ef}}{I_0}\times 100\%$$
Since $S$ is linear in $I$ we can apply the recorrection given in equation \ref{eq:Intensitycorrection} to obtain $S_f$, the true percentage of $I_0$ that is scattered forwards:
\begin{equation}
S_{f} = \frac{\alpha_t-\alpha_a}{\alpha_t}\left(\frac{1-e^{-\alpha_td}}{e^{-\alpha_ad}-e^{-\alpha_td}}\right)S_{mf} - S_{BGf} \label{eq:7}
\end{equation}
where $S_{BGf}$ is the background scattering signal measured in the forwards directions. Similarly for scattering backwards:
\begin{equation}
S_{b} = \frac{\alpha_t+\alpha_a}{\alpha_t}\left(\frac{1-e^{-\alpha_td}}{1-e^{-(\alpha_t+\alpha_a)d}}\right)S_{mb} - S_{BGb} \label{eq:8}
\end{equation}
where $S_{b}$ is the percentage of incident light scattered backwards, $S_{mb}$ is this percentage attenuated by absorption, and $S_{BGb}$ is the percentage scattered backwards in the background measurement. 

\subsection{Determining the Scattering Coefficient}
Finally, we must relate these to the total scattering coefficient, $\alpha_s$.  Combining equations \ref{eq:3} and \ref{eq:6} and similar equations for backscattering we find that the total scattering, $S = S_f + S_b$, is given by:
$$\frac{S}{100} = \frac{\alpha_{s}}{\alpha_t}\left(1-e^{-\alpha_td}\right)$$
Combining this with equations \ref{eq:7} and \ref{eq:8} we find:
\begin{eqnarray}
\frac{\alpha_t-\alpha_a}{\alpha_t}\left(\frac{1-e^{-\alpha_td}}{e^{-\alpha_ad}-e^{-\alpha_td}}\right)\frac{S_{mf}}{100} - \frac{S_{BGf}}{100} &+& \frac{\alpha_t+\alpha_a}{\alpha_t}\left(\frac{1-e^{-\alpha_td}}{1-e^{-(\alpha_t+\alpha_a)d}}\right)\frac{S_{mb}}{100} - \frac{S_{BGb}}{100}\nonumber\\
 &=& \frac{\alpha_s}{\alpha_t}\left(1-e^{-\alpha_td}\right) \label{eq:final}
\end{eqnarray}
Since $\alpha_a = \alpha_t - \alpha_s$ this equation has only one unknown ($\alpha_s$) and can be solved ($S_{mf}$, $S_{mb}$, $S_{BGf}$, $S_{BGb}$, $\alpha_t$ and $d$ are all measurable).  This must be done numerically, since $\alpha_s$ appears non-trivially on both sides.

\section{Results and Discussion}

\begin{figure}[p]
	\centering
		\includegraphics[width=10cm]{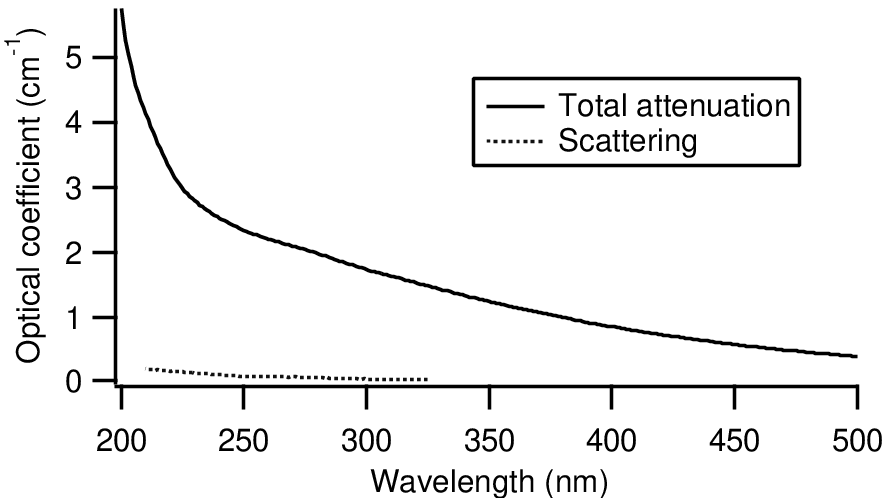}
	\caption{}
	\label{fig:ScatvsAbs}
\end{figure}

Figure \ref{fig:ScatvsAbs} shows the absorption coefficient for a $0.0025\%$ (by weight) solution of synthetic eumelanin over the visible and UV range.  It is typically broadband, and in excellent agreement with previously published absorption spectra of eumelanins \cite{Meredith04, Wolbarsht81, Nofsinger01, Krysciak85, Nofsinger99, OuYang04, Nofsinger01B, Nofsinger02}.  The measured scattering coefficient for the same solution is also shown, as a function of wavelength between 210nm and 325nm (calculated using equation \ref{eq:final}).  For wavelengths longer than 325nm the scattering coefficient was less than the minimum sensitivity of the instrument.  We expect that scattering will decrease at longer wavelengths; Rayleigh scattering, for particles with radii smaller than $\sim$ 50nm has a  $\lambda^{-4}$ dependance, and Mie scattering, for larger particles, is independant of wavelength. It is therefore reasonable to assume that the scattering coefficient is less than the measured values over the whole visible range.

\begin{figure}[p]
	\centering
		\includegraphics[width=10cm]{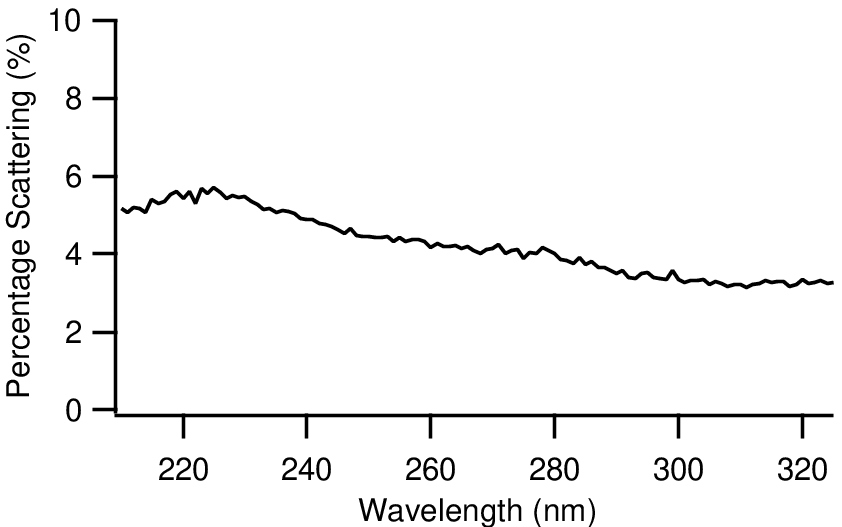}
	\caption{}
	\label{fig:PercScat}
\end{figure}

The percentage of the total attenuation due to scattering ($\alpha_s/\alpha_t\times100$) was calculated as a function of wavelength, and is plotted in figure \ref{fig:PercScat}.  It can be shown that the ratio of the coefficients is equivalent to the ratio of the intensities:
\begin{equation}
\frac{\alpha_s}{\alpha_t} = \frac{I_s}{I_s + I_a}
\end{equation}
where $I_a$ is the intensity of light lost due to absorption and $I_s$ is the intensity of light lost due to scattering.  Hence this quantity gives the percentage of the lost intensity that is due to scattering. It can be seen from figure \ref{fig:PercScat} that scattering contributes less than $6\%$ of the total loss at all wavelengths within the measured range.  This means that measured absorption spectra (total loss spectra) of eumelanin can be assumed to be primarily due to actual absorption, and used for interpretation of spectroscopic data without further manipulation.  This allows accurate determination of important quantities such as the radiative quantum yield of eumelanin \cite{Meredith04}.  This percentage is less than that measured by Vitkin et al. ($12\%$ at 580nm and $13.5\%$ at 633nm) and possibly indicates less aggregation in our more dilute solutions \cite{Vitkin94}.

\subsection{Prediction of Scattering Coefficient}
\begin{figure}[p]
	\centering
		\includegraphics[width=10cm]{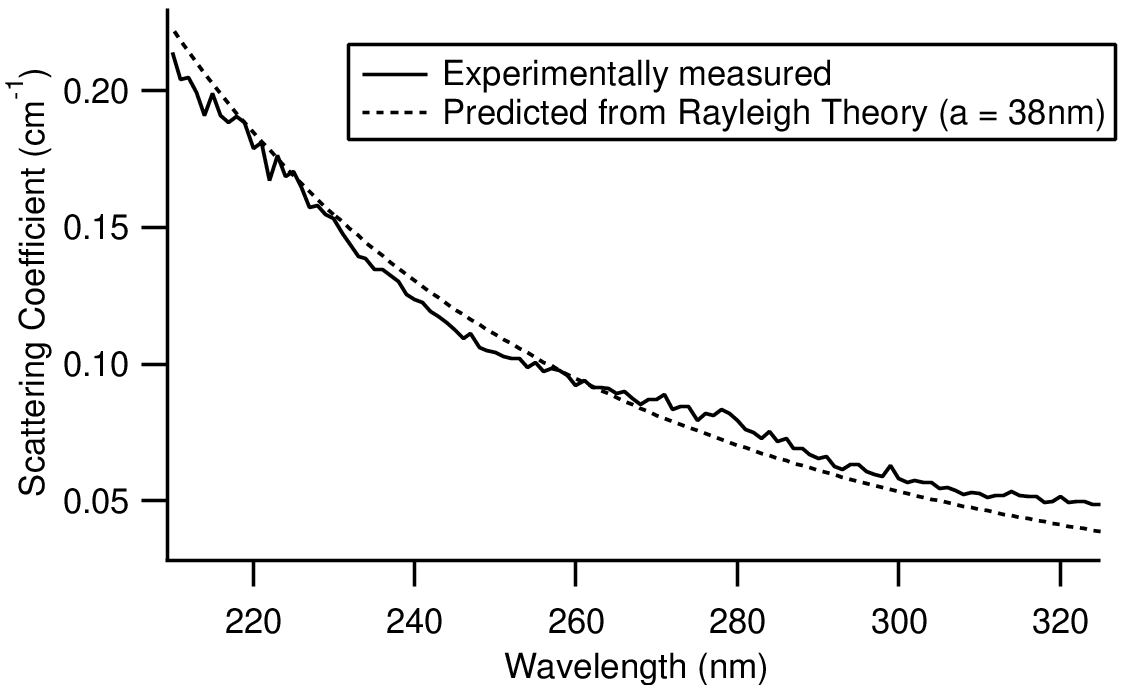}
	\caption{}
	\label{fig:ScatwithFit}
\end{figure}

The scattering coefficient appears to exhibit a dependancy upon the wavelength (figure \ref{fig:ScatwithFit}) which is suggestive of Rayleigh scattering, rather than Mie scattering (which is independant of wavelength).  Let us therefore determine whether the measured scattering coefficient is consistent with Rayleigh scattering alone (no Mie scattering).  As shown by Jackson \cite{Jackson99}, in the Rayligh limit (particles much smaller than the wavelength of the incident light), the scattering coefficient ($\alpha_s$) for dielectric spheres of radius $a$ with dielectric constant $\epsilon$ in a vacuum is given by:

$$\alpha_s = \frac{128\pi^5}{3}\frac{Na^6}{\lambda^4}\left|\frac{\epsilon-1}{\epsilon+2}\right|^2$$

where $\lambda$ is the wavelength of the illuminating light and $N$ is the number of spheres per unit volume.  This calculation can be repeated with the spheres in a solvent of dielectric constant $\epsilon_s$ to show that the scattering coefficient is then given by:

\begin{equation} \label{eq:Rayleigh_original}
\alpha_s = \frac{128\pi^5}{3}\frac{Na^6}{\lambda^4}\left|\frac{\epsilon-\epsilon_s}{\epsilon+2\epsilon_s}\right|^2
\end{equation}

Hence, knowing the way that the scattering coefficient depends upon the wavelength, we can estimate the size of the particles giving rise to scattering.  Unfortunately, it is nontrivial to apply this to melanins, since the structure of the fundamental particles is unknown.  This makes determining the number of particles per unit volume challenging.  Nevertheless, we can make some assumptions about the structure to determine an estimate of the particle size.

In the absence of a better structural model, it is a fair assumption that eumelanin monomers form approximately spherical particles.  The volume of each particle will be equal to the number of monomers per particle ($n_p$) multiplied by the `volume of a single monomer' ($V_m$) which can be estimated to be $1.2\times10^{-28}m^3$ \cite{Cheng94A, Cheng94B, Clancy01}.  Hence:

$$n_p = \frac{4}{3}\pi a^3 \frac{1}{V_m}$$

The molecular weight of a dihydroxyindole monomer is 149g/mol.  The molecular weight of an aggregate will therefore be $149n_p$g/mol.  Let $C$ be the concentration of our solution in weight percent, such that $C=2.5\times10^{-5}$ for a solution that is $0.0025\%$ eumelanin by weight.  Taking the density of the solvent (water) to be $1g/cm^3$, $1cm^3$ of solution will contain $C$ grams of eumelanin, or $C/(149n_p)$ moles of eumelanin aggregates.  The number of aggregates per $cm^3$ of solution will then be given by:

\begin{eqnarray*}
N &=& \frac{N_AC}{149n_p} \\
 &=& \frac{3N_ACV_m}{596\pi a^3}
\end{eqnarray*} 

where $N_A = 6.02214\times10^{23}$ is Avogadros number.  Applying this to Eq. (\ref{eq:Rayleigh_original}) we find:

\begin{equation} \label{eq:Rayleigh_melanin}
\alpha_s = \frac{32}{447}N_A\pi^4CV_m \frac{a^3}{\lambda^4}\left|\frac{\epsilon-\epsilon_s}{\epsilon+2\epsilon_s}\right|^2
\end{equation}

The dielectric constant for eumelanin ($\epsilon$) has been measured to be $\approx 2.72$ at optical frequencies (633nm) \cite{Kurtz86Dielectric, Kurtz86Book}.  The dielectric constant for water ($\epsilon_s$) is known to be $\approx 1.81$ at optical frequencies \cite{Pope97, Segelstein81}.  $V_m$ has been estimated to be $1.2\times10^{-28}m^3$, as discussed above.  Knowing these parameters we can fit the scattering coefficient vs wavelength curve by varying the particle size, $a$.  Although we have used several very rough assumptions about the structure of eumelanin, the particle radius is to the third power in the equation for the scattering coefficient. The scattering is therefore strongly dependant upon the particle size and it can be determined somewhat accurately from a measurement of scattering. 

This was done over the range 210nm to 325nm where accurate scattering data was available, as shown in figure \ref{fig:ScatwithFit}.  The best fit was found for a particle radius of $38\pm1$nm.  The good fit of the data to Rayleigh scattering theory suggests that we are in fact measuring scattering, and not another phenomemon (instrumental or otherwise).  This particle size is larger than that predicted by Cheng et al. \cite{Cheng94A, Cheng94B}, and possibly suggests that the protomolecules further aggregate.  Larger particles were measured by Vitkin et al., who report a particle size distribution for a similar sample preparation that has most particles with radii in the range 10-70nm \cite{Vitkin94}.  Hence an approximate particle size of 38nm is reasonable.

\section{Conclusion}
The integrated scattering of a eumelanin solution was measured as a function of incident wavelength, and found to contribute less than $6\%$ of the optical density between 210nm and 325nm.  This means that eumelanin absorption spectra can be interpreted as actual absorption with a high degree of confidence, and allows the calculation of many other optical spectroscopic quantities, such as the radiative quantum yield, without direct subtraction of scattering \cite{Meredith04}.  Hence, as long as eumelanin spectroscopic solutions are appropriately prepared and well solubilised, scattering is not a concern.  The scattering coefficient vs wavelength was found to fit Rayleigh Theory with a particle radius of $38\pm1$nm.  This is a larger estimate of the fundamental particle size than those previously reported from X-ray scattering and microscopy studies \cite{Clancy01, Cheng94A, Cheng94B}, and perhaps indicates that in our samples the fundamental particles have aggregated.  This is consistent with other optical studies \cite{Vitkin94}.  Knowing the physical structure of eumelanin particles is essential for interpretation of spectroscopic results, and therefore for understanding the de-excitation pathways in eumelanin and its biological functionality.

\section*{Acknowledgements}
This work has been supported by the Australian Research Council, the UQ Centre for Biophotonics and Laser Science, and the University of Queensland (RIF scheme).


\section*{Figure Captions}
\begin{enumerate}
\item a) Geometry for $100\%$ transmission standard.  b) Geometry to collect forward scattered light  c) Geometry to collect backwards scattered light.
\item Cuvette geometry.
\item Total attenuation and scattering coefficients for a $0.0025\%$ (by weight) solution of synthetic eumelanin.
\item The scattering coefficient (as plotted in Figure \ref{fig:ScatvsAbs}) as a percentage of the total attenuation coefficient for the same solution.  We see that even over this short wavelength range where scattering should be most significant, it contributes less than $6\%$ of the total attenuation.
\item The eumelanin scattering coefficient, with the predicted Rayleigh scattering coefficient (from Eq. (\ref{eq:Rayleigh_melanin})).  The best fit (plotted above) was obtained with a particle radius of 38nm.
\end{enumerate}

\clearpage
\section*{Figures}

\clearpage

\clearpage

\clearpage

\clearpage

\end{document}